# Three-dimensional Insight on the Evolution of a Supramolecular Preorganization Complex to Hollow-Structure Carbon Nitride

*Wu Wang, Zongzhao Sun, Jianghe Feng, Juan Cui, Limin Huang* and Jiaqing He**

Dr. W. Wang, Dr. J.H. Feng, J. Cui, Prof. J.Q. He
Department of Physics, Southern University of Science and Technology, Shenzhen, Guangdong, 518055, China
E-mail: hejq@sustech.edu.cn

Z.Z. Sun, Prof. L.M. Huang
Department of Chemistry, Southern University of Science and Technology, Shenzhen, Guangdong, 518055, China
E-mail: huanglm@sustech.edu.cn

**Abstract**: The supramolecular preorganization approach can be applied to effectively fabricate various morphologies of graphitic carbon nitride (g-CN) with improved photocatalytic activity, while a comprehensive understanding for the morphology evolution from supramolecular aggregates to g-CNs is lacking. Herein, 3D characterizations from electron tomography provide a fundamental insight on the evolution from rod-like melamine–cyanuric acid (MC) complex to hollow-structure g-CN in the thermal polycondensation process. The internal region and a group of surfaces of the rod-like complex initially underwent polycondensation, while the other two groups of surfaces with ~100 nm thickness almost unchanged. With the temperature reached to 550 °C, the hollow-structure g-CN eventually formed due to most of the internal matter vanishing, and voids arose in the previously unaffected surfaces (edges), resulting in a porous shell structure. Quantitative electron tomography indicates that the key of structure evolution is the differentiated condensation polymerization between edges and inner region of the rod-like MC complex, which is ascribed to a higher dense of surfaces and a lower dense of inner matter with loose, defective orignization.

# 1. Introduction

Graphitic carbon nitride (g-CN), as a metal-free polymeric semiconductor, has attracted widespread attention in the last decade owing to its low cost, good thermal and chemical stability, and an appealing electronic structure for visible-light response.[1–4] These unique features make g-CN a promising material in the field of photocatalysis, including applications in the water splitting for $H_2$ and $O_2$, the pollutant degradation, and the reduction of $CO_2$ to renewable hydrocarbon fuels.[5–7] To promote this material for practical photocatalytic applications, designing micro-nanostructures of g-CN is considered to be an effective approach for improving the specific surface area, increasing amount of active sites and enhancing utilization of visible light, in which specific morphologies of g-CN often imply high photocatalytic activity.[6–11]

At present, the supramolecular preorganization complex of melamine and cyanuric acid or derivatives therefrom, in which molecules joined by non-covalent bonds assemble into stable aggregates under equilibrium conditions, is considered to be a promising method to fabricate specially-shaped g-CN materials.[12–16] For instance, Jun et al.,[15] demonstrated the formation of hollow g-CN spheres from thermal polycondensation of flower-like, spherical, supramolecular aggregate using the melamine and cyanuric acid as precursors dissolved in dimethyl sulfoxide. Shalom et al.,[14] used the melamine–cyanuric acid complex in ethanol as the precursor to further fabricate the pancake-like g-CN hollow replicas with thermal treatment. Moreover, a well-defined tubular g-CN was obtained from the pyrolysis of the hexagonal rod-like supramolecule with melamine and phosphorous acid precursors dissolved in water.[16] In most cases, the supramolecular preorganization approach includes two steps for preparing specially-shaped g-CNs: i) the formation of a supramolecular complex from monomer self-assembly and ii) thermal polycondensation of resultant supramolecular aggregates,[15] in which the morphology of the supramolecular complex directly relates to the structure of assembled molecules. The thermal polycondensation procedure is of particular importance for producing the desired hollow-structure g-CNs, which often possess larger surface area, faster mass transfer and enhanced light-harvesting, and therefore significantly improved activity compared to bulk g-CNs in the photocatalysis.[14–18]

However, understanding of the morphology evolution from supramolecular aggregates to g-CNs during the thermal polycondensation process is still limited, as previous investigations were mainly based on two-dimensional (2D) micrographs[12,14,16] and thus insufficient to reflect their genuine three-dimensional (3D) architectures. Electron tomography, which reconstructs the 3D morphologies from a series of 2D projections in the transmission electron microscopy (TEM), has demonstrated the capabilities to resolve intricate 3D structures among a variety of materials,[19–21] such as diverse morphologies of $CeO_2$ nanoparticles,[22] mesoporous silica/carbon,[23,24] zeolites[25–27] and supported nanocatalysts[9,28,29]. Herein, we employ electron tomography to systematically investigate the 3D morphology evolution of the melamine–cyanuric acid (MC) complex to the hollow-structure g-CN during the thermal polycondensation at different temperatures. The MC supramolecular complex prepared by the method used in our previous work[30] has a well-defined, hexagonal rod shape, and three intermediate morphologies are determined from the polycondensation at 350 °C and 450 °C. Electron tomography, in combination with image analysis, unambiguously reveals 3D structures and inner features of rod-like MC supramolecule, intermediate complexes and the resultant hollow-structure g-CN from 550 °C polycondensation. Inhomogeneous denseness of the edges and inner region has been found in the hexagonal rod-like supramolecule and is considered the key to generate the hollow-structure g-CN during the thermal treatment. Moreover, a 3D model combined with chemical structures has been proposed for further clarifying the morphology evolution of the rod-like MC complex to the hollow-structure g-CN during thermal polycondensation.

## 2. Results and Discussion

Figure 1 shows typical scanning transmission electron microscopy (STEM) images of the as-synthesized MC supramolecular complex, products of the complex calcined at 350 °C and 450 °C, and the resultant g-CN from heating treatment at 550 °C. The as-synthesized MC complex had a well-defined rod shape with nearly uniform contrast (Figure 1a), while a few cracks along the longitudinal direction seem to arise in the rod-like complex after 350 °C calcination (Figure 1b). Moreover, two typical structures were found in the complex under the 450 °C polycondensation (Figure 1c and 1d), where the number of voids increased significantly in the center parts due to the accompanied substance and mass loss[15,31]. In the thermal polycondensation at 550 °C, most of the solids disappeared and some bubble-like features remained in the center, resulting in a hollow structure with a thin shell (Figure 1f).

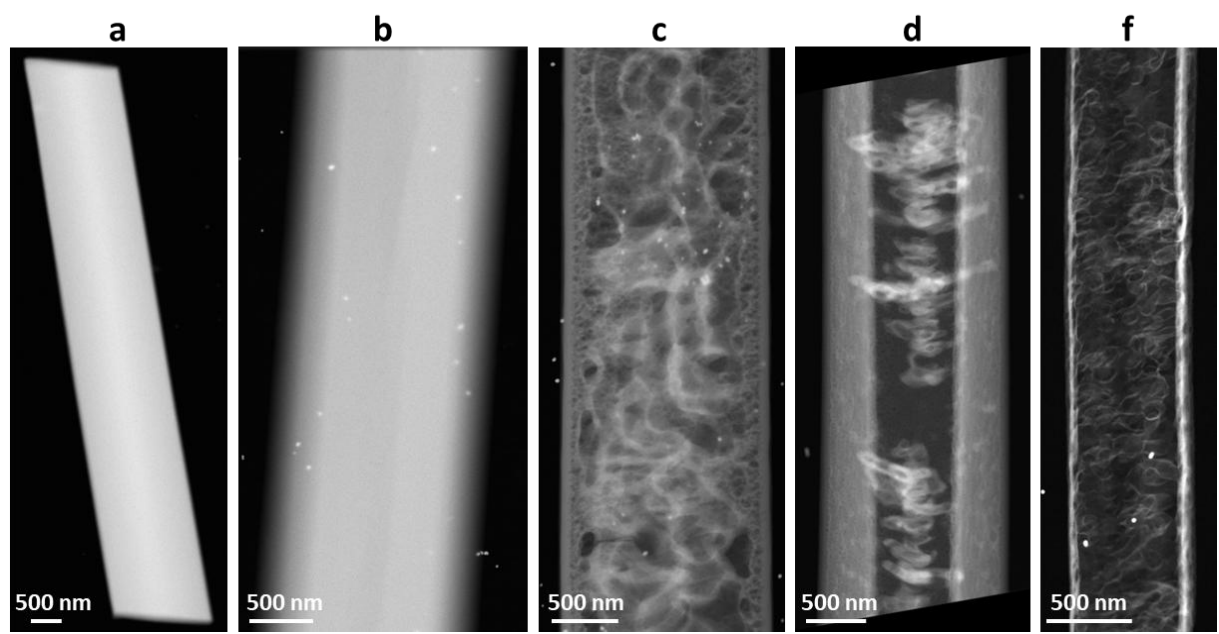

**Figure 1.** STEM images from tilt-series of (a) as-synthesized MC supramolecular complex, (b) the complex after 350 °C polycondensation, (c, d) typical structures of the complex at 450 °C polycondensation and (f) the resultant g-CN from 550 °C polycondensation. The bright particles in the images are gold particles deposited on the grid as tracking markers for electron tomography.

However, the above results are insufficient to comprehensively understand the structure evolution of the MC supramolecular complex during polycondensation, as STEM imaging is a 2D projection of the 3D object and remains ambiguous for investigating intricate 3D structures, especially for structures with inner pores and voids in our case. 3D characterizations of these structures evolved from the MC complex to the hollow structure g-CN were performed by electron tomography (Figure 2), where the structure obtained in 350 °C polycondensation and two typical structures during 450 °C polycondensation are denoted as "intermediate-1", "intermediate-2" and "intermediate-3" for classification, respectively. The surface rendering and the cross-sectional slices parallel and perpendicular to the longitudinal axis were applied to interpret the 3D reconstructions (a combined view of the surface rendering and slices is given in supporting information Figure S1 and Movie M1). For the as-prepared MC complex, there is a well-defined hexagonal structure for the rod-like particle (Figure 2a), consistent with previous researches[16,30]. The inhomogeneous contrast of 3D reconstruction is clearly revealed in both longitudinal and lateral slices, indicating the brighter edges and the darker inner region, which will be discussed in detail later. In the case of intermediate-1, cracks and a few voids with lengths from tens to hundreds of nanometers are determined unambiguously in regions close to the edges from the 3D characterization, while the inner part is almost unchanged (Figure 2b). The structure of the MC complex suffers a severe evolution with increased calcination temperature, as shown in Figure 1c, 1d and 2c, 2d. In the intermediate-2, larger voids with a size of hundreds of nanometers are found in the inner region while lots of smaller voids occur in the edges of the particle, resulting in a non-uniform pore structure (Figure 2c). This could indicate the structural difference between the edge and inner regions of the raw MC complex and therefore an inhomogeneous polymerization during condensation. Most importantly, the differentiated polymerization of the hexagonal edges is also revealed from lateral views of the 3D surface rendering and cross-section slices (Figure 2c (ii) and (iv)), in which the upper and lower surfaces show high porosity due to the mass loss, while the other two groups of surfaces with ~100 nm thickness remain a nearly-solid state. The differentiated polymerization of edges still retains in the intermediate-3 structure, as shown in Figure 2d. But the voids clearly expand in the interior space, leaving a discontinuous, porous framework. The hollow structure was achieved eventually with the calcination temperature reaching 550 °C, which is the typical temperature used for fabricating g-CNs in the supramolecular preorganization approach[14–16]. Figure 2e indicates that the residual framework nearly vanished in the inner region and a lot of voids arose in the edges that were unaffected in the previous polycondensation, forming a porous shell structure or even a double-layered shell (Figure 2e (iii) and (iv)).

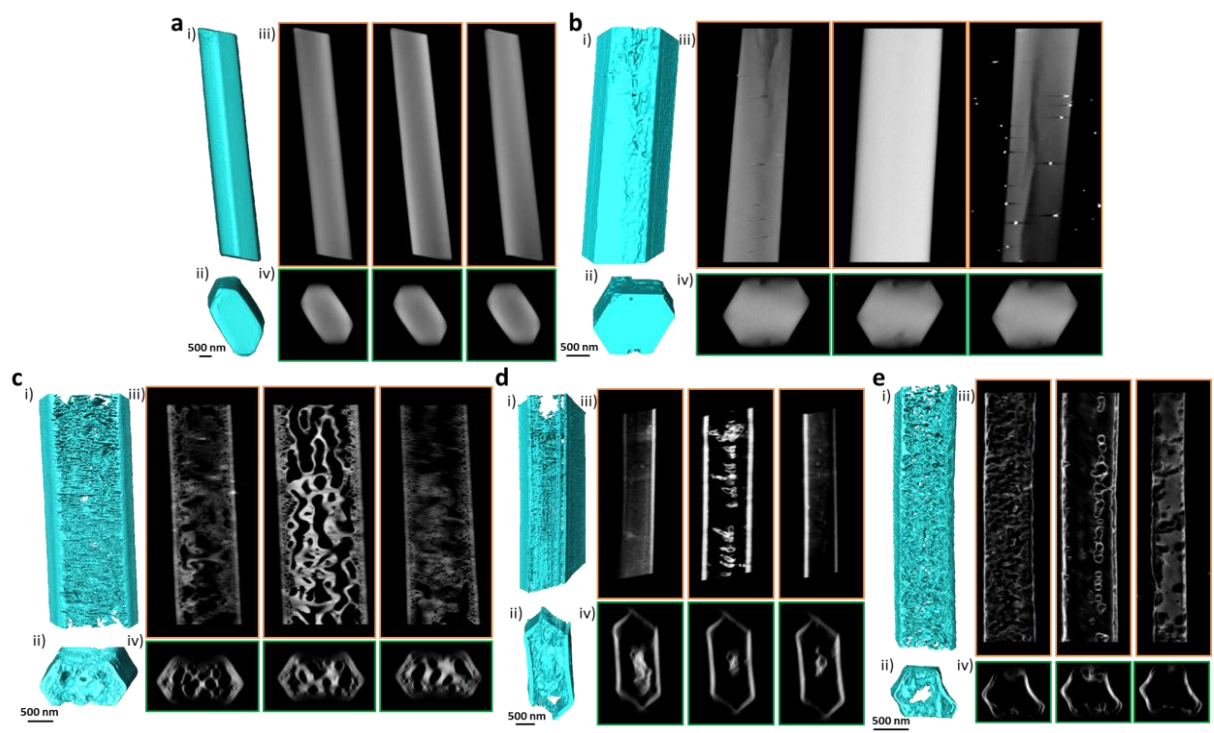

**Figure 2.** Electron tomography analysis of (a) rod-like supramolecule, (b–d) intermediate-1, 2, 3 and (f) the hollow structure g-CN with (i) longitudinal and (ii) lateral views of the 3D surface rendering and cross-sectional slices (iii) parallel and (iv) perpendicular to the long axis. The bright dots in the image (b) are related to gold particles deposited on the grid as tracking markers for electron tomography and the 3D views of surface rendering for the five structures are given in supporting information Movie M2-6.

Electron tomography not only can qualitatively visualize 3D structures, but also enable to provide quantitative information.[9,23,29,32,33] The porosity of whole structures was calculated from segmented reconstructions, as shown in supporting information Figure S2a. During the structure evolution in thermal polycondensation, the porosity increased rapidly from the as-prepared complex and intermediate-1 to intermediate-2 and intermediate-3, and slightly enhanced in the obtained hollow structure, which agrees with the mass loss from the thermogravimetric analysis (TGA) measurement of the as-prepared complex (supporting information Figure S2b). Moreover, the differentiated polymerization in the morphology evolution was further quantified from the porosity of edges and inner regions in the reconstructions by the statistics of void and solid volumes within ~100 averaged slices between each group of opposite edges (the scheme is shown in Figure 3a and the corresponding description is included in the Experimental Section). The porosity of the inner region roughly increased in the structure evolution from a rod-like complex to the hollow structure during the polycondensation process (Figure 3b–d). Irregular porosity was found in the intermediate structures, probably resulting from the inhomogeneous polymerization of internal matter. The central framework of the intermediate-3 is well-determined from the porosity statistical analysis and subsequently vanishes in the hollow structure, giving rise to higher porosity of the inner region compared with that of the edges. More importantly, the differentiated polycondensation of edges in the complex is unambiguously revealed from the quantification analysis. The group of edges (Figure 3b) evolved the polycondensation together with the inner region, showing a continuously increasing porosity, while the other two groups of edges were almost unchanged until the temperature reached to 550 °C, indicated by the variation of their porosity summarized in Figure 3c and 3d.

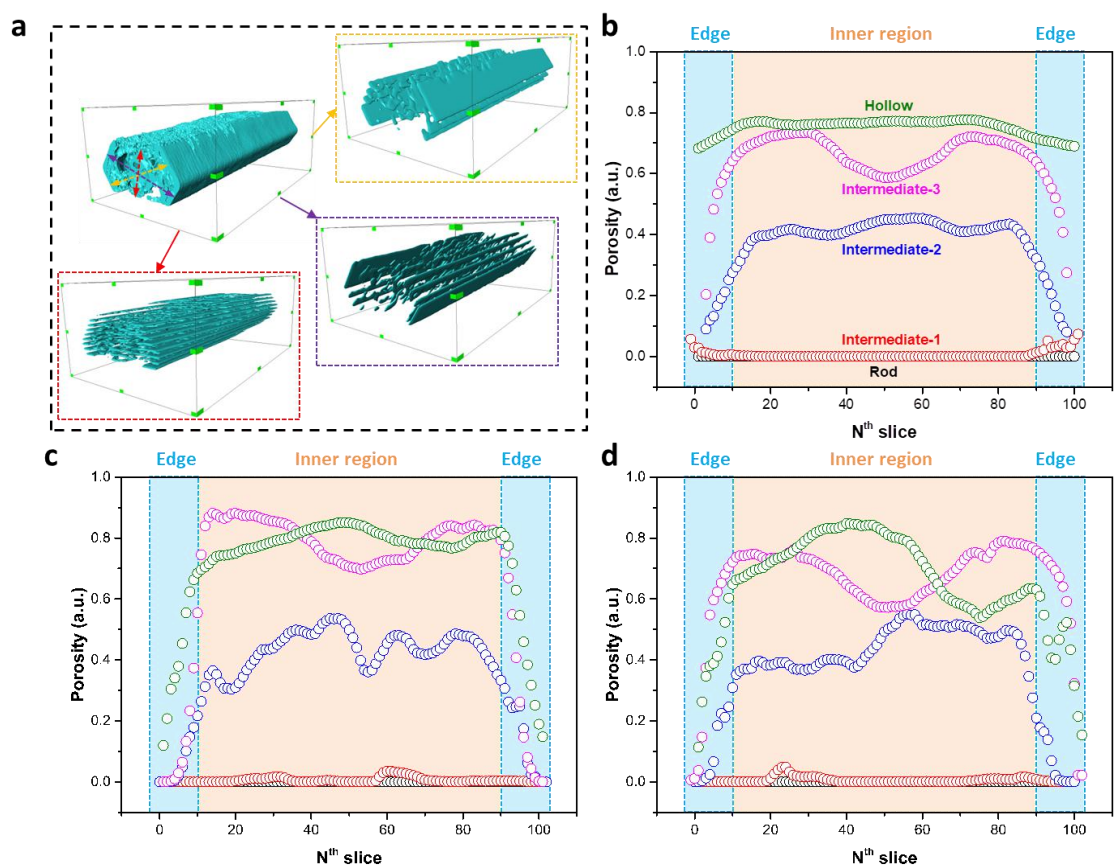

**Figure 3.** (a) Scheme of segmented reconstructions disassembled along the normal of edges averaged to ~100 slices and (b–d) the porosity of averaged slices across opposite edges in the rod-like supramolecule, intermediate structures, and hollow structure.

The aforementioned 3D characterizations have demonstrated the inhomogeneous polycondensation of edges and inner region in the structure evolution from a rod-like complex to the hollow structure. The observed phenomenon is further elucidated by the 3D density of reconstruction volume from the rod-like complex, which is plotted by the direct volume rendering that is an intuitive method for visualizing 3D scalar fields based on the assumed absorption light of each point in the 3D volume[34]. Figure 4a shows a higher dense of surfaces (edges) and a lower dense of inner region in the rod-like complex, and the cross-section slice (Figure 4b) further points out the inhomogeneous density of hexagonal edges, in which a group of surfaces has a similar dense to the inner region and the other two groups of surfaces with ~100 nm width possess an enhanced denseness. This is also confirmed by the intensity profile across opposite edges in the raw slice (Figure 4c and 4d).

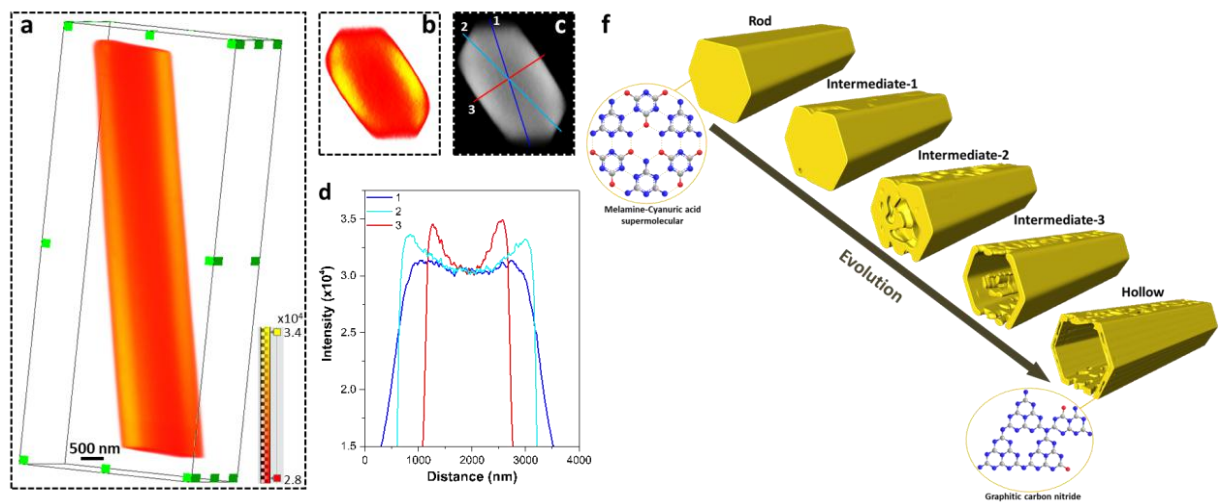

**Figure 4.** (a) The 3D density plot using the volume rendering approach, (b) the density of the 2D cross-section slice, and (c) the raw slice with gray-scale of the rod-like complex, (d) the intensity profile of highlighted lines across opposite edges in the hexagonal slice (c), (f) scheme of 3D models from the melamine–cyanuric acid complex to the hollow structure g-CN during thermal polycondensation. (The gray, blue and red balls represent C, N and O atoms, respectively, and the dash lines denote the hydrogen bonds in the chemical structures of the MC supramolecule and g-CN. H atoms are neglected for simplicity)

Moreover, the observation on the cross-section of the rod-like complex prepared by the ultra-microtome indicates a loose structure with defects in the center (supporting information Figure S3a and S3b) and thus a lower dense of the inner region, probably due to quickly crystallizing after the formation of the supramolecular complex from the self-assembly of monomers. The growth along the [001] direction of the rod-like complex has been revealed from electron diffraction (supporting information Figure S3c) of this cross-section, indicating the possible defective structures in the (110) planes. Rietveld refinements for X-ray diffraction (XRD) pattern of the rod-like compound further confirm this point, as shown in supporting information Figure S3d and S3e. The mismatch between the experimental XRD pattern and simulated XRD pattern from the MC ($C_{12}H_{18}N_{18}O_6$) model with the monoclinic *I2/m* structure[35] is reduced obviously by additionally applying the structure with defective (110) planes, which is reflected on the decreased reliability factors $R_p$ and $wR_p$. Based on the 3D characterizations, the models for the structure evolution from the rod-like complex to the hollow structure during thermal polycondensation are proposed in Figure 4f. During the thermal process, internal matter and a group of surfaces of the rod-like MC complex evolve the polymerization together at 350 °C and 450 °C, and it eventually forms the hollow structure g-CN at 550 °C with a porous shell from the polycondensation of the other two groups of surfaces.

XRD and X-ray photoelectron spectroscopy (XPS) reveal the chemical structures of the rod-like complex and the hollow structure product, as shown in Figure 5. The hexagonal rod-like MC precursor is prepared through the molecular self-assembly between melamine and cyanuric acid that is from the *in situ* hydrolysis of melamine under the hydrothermal[16,30]. The as-prepared MC supramolecular complex has a highly ordered interplanar stacking reflected from peaks at 10.9°, 11.9° and 22.0° in the XRD pattern (Figure 5a), while these peaks shift slightly to lower angles and become weaker for the complex calcined at 350 °C, indicating the partial hydrogen bonds fracture among melamine and cyanuric acid. With the calcination temperatures increased to 450 °C, all peaks of the MC complex disappear completely and two arising peaks at 13.2° and 27.3° are related to in-plane ordering and interlayer stacking of tri-s-triazine units[1], indicating the formation of g-CN. Further condensation of in-plane tri-s-triazine units under the 550 °C calcination is also demonstrated from the enhanced intensity of the (100) peak.

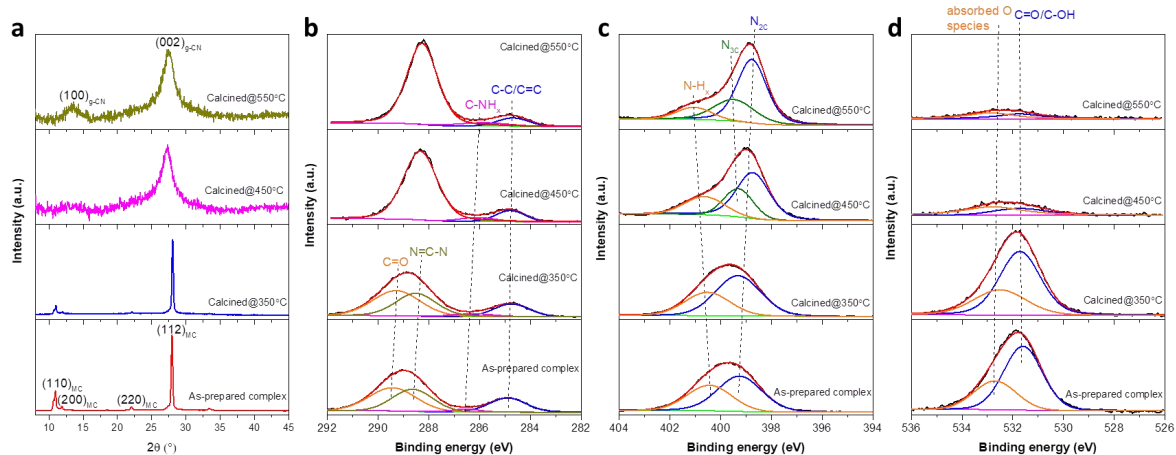

**Figure 5.** (a) XRD patterns and XPS spectra of (b) C 1s, (b) N 1s and (c) O 1s of the as-prepared MC supramolecule and its products from calcination at 350 °C, 450 °C, and 550 °C.

The chemical environments and electronic structures of the different stages were further indicated by XPS. In the C 1s spectra (Figure 5b), the peak at 289.5 eV corresponding to C=O bonds vanish and the peaks of N=C-N and C-NH$_x$ shift to lower energies as the temperature reaches 450 °C and 550 °C, suggesting the decomposing of cyanuric acid and the gradual release of the oxygen species for the weaker electron-withdrawing group, thus resulting in higher electron cloud density around N=C-N and C-NH$_x$. This phenomenon is also confirmed from the shift of the N$_{2C}$ peak in the N 1s spectra (Figure 5c). In addition, the statistical peak areas of N-H$_x$ decrease due to the releasing of NH$_3$ and condensation, and a new peak arises at 399.4 eV for N$_{3C}$ from 450 °C and 550 °C calcination, indicating the formation of bridged N and the core N of melon. The decreased peak areas around 532.7 eV in the O 1s spectra (Figure 5d) further demonstrate the releasing of oxygen species in thermal treatment. The remained peak at 531.8 eV regarding C=O/C-OH indicates the oxygen dopants in the g-CN obtained at 550 °C condensation. Therefore, the chemical structures of the MC supramolecule and obtained g-CN are logically proposed in Figure 4f. Moreover, based on the aforementioned investigations, the whole chemical structure evolution of the MC supramolecule to the hollow structure g-CN is proposed systematically in supporting information Figure S4. The weaker hydrogen bonds of the MC supramolecule start to break at 350 °C calcination. The part of dissociative monomers could polymerize to form linear polymers, such as melon and melamine chain at around 450 °C with the release of $NH_3$, $CO_2$ and $H_2O$ gas. These linear polymers further crosslink to the net structure at the early stage of 550 °C and complete the thermal-polymerization after incubation with prolonged time, in which the unstable triazine-based motifs (poly(triazine imide)) could decompose and transfer to the more stable tri-s-triazine-based g-CN.[36–39]

## 3. Conclusion

Electron tomographic reconstructions provide the 3D insight to unambiguously reveal the morphology evolution of the rod-like MC complex to the hollow structure g-CN in the thermal polycondensation of the supramolecular preorganization approach. The observed intermediate structures indicate that the internal region, together with a group of surfaces of the MC complex, initially undergo the polycondensation, thus resulting in a porous structure, while other two groups of surfaces with ~100 nm thickness almost remain in the solid state. With the thermal treatment temperature reached to 550 °C, a hollow structure is eventually obtained and voids arise in previously unaffected edges, forming a porous shell structure or even a double-layered shell. The morphology evolution of the rod-like MC complex to the hollow structure g-CN originates from the differentiated condensation polymerization between the edges and inner region of the as-prepared MC complex, which is ascribed to the denseness variation, a higher dense of surfaces (edges) and a lower dense of inner matter. This could result from the defective structures in the (110) in-plane stacking at the center part of the rod-like complex. Our work offers a fundamental insight into the morphology evolution in thermal polycondensation of the supramolecular preorganization approach for preparing specific morphologies of g-CNs. The 3D characterizations from electron tomography have the capability to explore diverse morphologies of g-CNs in various synthesis methods, which is beneficial for designing and/or optimizing their micro-nanostructures and therefore achieving high photocatalytic properties.

## 4. Experimental Section

*Synthesis and thermal treatment*: The preparation of the rod-like melamine-cyanuric (MC) supramolecule followed the method used in our previous work[30]. Briefly, melamine was firstly dissolved in deionized water at 70 °C and then transferred to a cold-water bath with ultrasonication for the recrystallization of melamine. The obtained suspension was put into a Teflon-lined stainless autoclave and heated at 200 °C for 12 h in an oven. After cooling down, the white products were collected by filtration and dried at 60 °C for further usage. The heating treatment of as-prepared supramolecular complex was performed in an alumina crucible with a cover by thermally decomposing at 350 °C for 1h, 450 °C for 1h and 550 °C for 3 h, respectively, under the $N_2$ flow with a ramp rate of 5 °C/min.

*Electron tomography and 3D image analysis*: The powder of MC complex and products of the complex at from 350 °C, 450 °C and 550 °C thermal polycondensation were directly dispersed on 100 mesh carbon coated Cu grids, where 20 nm colloidal Au particles had been deposited beforehand. The tilt-series images over a tilt range of around ±70º with a 2° increment were acquired at high-angle annular dark-field (HAADF)-scanning transmission electron microscopy (STEM) mode using a Fischione 2020 tomography holder in a FEI Talos F200X microscope operated at 200 kV. The tilt series were aligned in IMOD[40] using the Au particles as fiducial markers with an achieved mean residual error of 0.47-0.49 pixels and further reconstructed by the simultaneous iterative reconstruction technique (SIRT) in the Inspect3D v4.4 (Thermo Fisher Scientific). The resultant tomograms had a final voxel size of 2.15, 3.03 or 6.07 nm, depending on the magnification of the original STEM images.

The reconstructed tomograms were input into the Fiji software[41] for image processing, where the median filter and the CLAHE plugin[42] with respect to the local contrast enhancement had been applied to reduce noise and get well-separated image intensities for solid and void parts to facilitate segmentation. Further visualization and analysis were performed in Avizo v9.0 (Thermo Fisher Scientific). The binary 3D reconstruction volumes were obtained using the global threshold and further separated into three parts: vacuum, void and solid by the hole filling approach implemented in Avizo that is to separate internal voids and vacuum around the investigated objects. The aforementioned image processing was illustrated in supporting information Figure S5. The segmented reconstructions were visualized by the 3D surface rendering and volumes of solid and void were calculated using the label analysis module to obtain the porosity of investigated specimen. Moreover, inner features of investigated particles were displayed by the longitudinal and lateral cross-section slices. To quantify the porosity of edges and inner region in the reconstructions, orientation of cross-sectional slices was adjusted to parallel to the hexagonal edges and then the segmented reconstructions were averaged along the normal of edges to generate ~100 slices between opposite edges, in which volumes of solid and void in the averaged slices were measured by the material statistics module.

*Ultra-microtomy and transmission electron microscopy*: The powder of rod-like MC supramolecular complex was embedded in a polyethylene BEEM capsule using an epoxy resin (Pon 812). A Leica ultra-microtome (UC7) with a diamond knife (DU4530) was used to prepare a nominal 100 nm thickness slices with a cutting velocity of 2 mm/s and distilled water as trough liquid at room temperature. Then the microtomed slices were transferred to a carbon coated Cu grid and analyzed in a FEI Talos F200X microscope operated at 200 kV, where the selected area electron diffraction (SAED) for the cross-section of rod was acquired with a dose rate of 0.1 e$^-$/Å$^2$/s for 20s.

*X-ray diffraction (XRD) and refinement*: The powder of investigated specimen were grinded to refine the particle size for XRD measurement, which was performed on a Rigaku smartlab system at 45 kV and 200 mA with Cu Kα radiation ($L_{K\alpha1}$ = 1.540593 Å, $L_{K\alpha2}$ = 1.544414 Å), at a scan rate of 2°/min and a step size of 0.020°. For the refinement of XRD pattern, the crystal structure parameters of MC supramolecule were adopted from reference [35] and the defects were further introduced to the (110) planes for generating a defective structure, by changing the distance between melamine and cyanuric acid molecules and breaking manually some of the hydrogen bonds. Rietveld refinements of the powder XRD pattern were performed using the program Jana 2006, with the original structure and the defective structure.

*X-ray photoelectron spectroscopy (XPS)*: The surface elements properties were analyzed by a XPS (Kratos XSAM800 spectrometer, USA) with Al K X-rays radiation operated at 300 W. The C 1 s level at 284.8 eV as an internal standard was used to correct the shift of binding energy. The spectra were fitted using a nonlinear least square fitting program (XPSPEAK41) with a Shirley background.